\documentclass[lettersize,journal]{IEEETran}
\usepackage[hidelinks, breaklinks]{hyperref}
\usepackage{graphicx,wrapfig,float,epstopdf}
\usepackage{subcaption}
\usepackage{amsmath,amsfonts,amsthm,amssymb,txfonts,gensymb,siunitx}
\usepackage{soul,color}
\usepackage[capitalise]{cleveref}
\graphicspath{{./figures/}} 
\usepackage{caption}
\usepackage{tikz}
\DeclareRobustCommand\full  {\tikz[baseline=-0.6ex]\draw[thick] (0,0)--(0.5,0);}

\DeclareRobustCommand\dashed{\tikz[baseline=-0.6ex]\draw[thick,dashed] (0,0)--(0.54,0);}

\usepackage{url}
\usepackage{breakurl}

\usepackage{authblk}

\providecommand{\keywords}[1]
{
  \small	
  \textbf{\textit{Keywords ---}} #1
}
\usepackage{natbib}
\setcitestyle{authoryear,open={(},close={)}} 

\usepackage{setspace}
\begin{document}

\title{Effects of eco-driving on energy consumption and battery degradation for electric vehicles at signalized intersections}

\author{Yongqiang Wang\thanks{Corresponding author: Yongqiang Wang, Email: sailor@udel.edu/yqwangjohn@gmail.com, Address: Department of Mechanical Engineering, 126 Spencer Lab, Newark, DE 19716}, Suresh G. Advani, Ajay K. Prasad}
\affil{Department of Mechanical Engineering, University of Delaware, Newark, DE, 19716, USA}

\markboth{arxiv submission, 10/1/2024}{arxiv submission, 10/1/2024}
\maketitle

\begin{abstract}
	Eco-driving has been shown to reduce energy consumption for electric vehicles (EVs). Such strategies can also be implemented to both reduce energy consumption and improve battery lifetime. This study considers the eco-driving of a connected electric vehicle equipped with vehicle-to-infrastructure (V2I) communication passing through two signalized intersections. Dynamic programming is employed to construct an eco-driving algorithm that incorporates a battery degradation model in addition to minimizing energy consumption to optimize the vehicle's speed trajectory while transiting the control zone. A parametric study is conducted for various signal timings and distances between the two intersections. It is found that eco-driving can provide up to 49\% in cost benefits over regular driving due to energy savings and improved battery life which could boost consumers' interests on EVs. This study also considered different battery capacity decay rates based on battery chemistry. Although a higher decay rate affects the optimal speed trajectories only slightly, it amplifies the benefits of eco-driving on battery life. Two battery sizes were also studied to show that the larger battery is associated with a drastically increased lifetime, thus creating opportunities for electric vehicles in other applications such as vehicle-to-grid (V2G) integration. Field tests were also conducted using a simplified rule-based version of the eco-driving algorithm implemented as a phone app which issues audio speed recommendations to the driver. The field test results were promising and validated the results from simulations. The phone app implementation is convenient and could facilitate broader adoption and widespread use of eco-driving which helps to improve transportation efficiency and protect the environment.
\end{abstract}


\section{Introduction}
Eco-driving is a technique used to reduce energy consumption of vehicles by optimizing velocity trajectories in different driving scenarios. It could be achieved either by driver training or with in-vehicle driver assistance devices by which drivers can adapt their driving behavior to improve fuel economy \citep{corridor, driving_assitant, train_bus_drivers}. However, simple approaches like minimizing deceleration levels cannot guarantee improvements without a detailed model of fuel consumption \citep{v2i}. Moreover, the effectiveness of driver training tends to fade over time and in-vehicle devices provide more consistent benefits \citep{review}. In-vehicle devices have become more promising due to recent advancements in connected and automated vehicles technologies which provide essential vehicle-to-infrastructure (V2I) information needed to implement eco-driving. 

An eco-approach and departure (EAD) system was tested in \citep{ead} which showed a 6\% saving in energy consumption and various reductions in air pollutant emissions. Dedicated short range communication (DSRC) was used to transmit signal phase and timing (SPaT) information to drivers who then modified their driving behavior accordingly to avoid unnecessary acceleration and deceleration. An advanced driver assistance system (ADAS) based on vehicle-to-cloud (V2C) communication was implemented in \citep{v2c} to provide advisory speed for a driver merging onto a ramp. The computation was carried out in the cloud and the test results proved the effectiveness of the system despite communication delays and packet losses. Uncertainty in the signal timing was also modeled in \citep{ERD, uncertain_traffic_signal_timing} using a data-driven method in which eco-driving was formulated as a robust optimization problem. The results showed a 40\% reduction in fuel consumption while maintaining similar arrival times compared to a modified intelligent driver model (IDM). Similarly, a reduction of around 30-50\% in energy consumption was observed in \citep{bus_with_stops} which studied speed optimization for electric buses travelling along signalized arterials with bus stops. An eco-speed control system was field tested in \citep{eco_speed} and showed a reduction of 17.4\% in fuel consumption and 8.4\% in travel time. A parameterized eco-driving algorithm designed for two signalized intersections was tested in \citep{eco_multi_signal} which showed higher fuel savings compared to algorithms designed only for a single intersection. An eco-driving scenario with wireless-charging around signalized intersections was studied in \citep{eco_with_wireless_charging} which showed increased driving range and reduced cost. A detailed powertrain model was used in \citep{eco_powertrain} to show that extremely low state-of-charge (SOC) could significantly affect the acceleration performance and efficiency of electric vehicles (EVs) while ambient temperature had little effect. Optimal depth-of-discharge (DOD) was considered in \citep{eco_hybrid_ebus} for a connected hybrid electric bus to improve fuel economy and reduce battery degradation. A predictive cruise controller was implemented for electric vehicles which showed a 7\% saving in energy consumption and 30\% increase in battery life \citep{cruise_control_battery_lifetime}. A similar study on electric trucks that considered highway topography and traffic showed a 5\% reduction in energy consumption and 100\% increase in battery life \citep{ev_trucks}.

Although eco-driving is not currently widespread, the benefits of eco-driving increase with market penetration and hence it is essential to seek higher penetration rates in the near future \citep{eco_approaching_partial, market_pen_cav}. A case where the vehicle is neither connected nor automated was studied in \citep{preceding_vehicle_non_connected} and a 4\% energy saving was observed compared with a conventional car-following strategy. The likelihood of passing an intersection was modeled as a stochastic event based on historical data in \citep{data_driven_eco} to advise the driver to slow down or accelerate to pass. The results showed that eco-driving is promising even without the widespread implementation of connected or automated vehicle technologies.

Despite dramatic improvements in capacity and performance of Lithium-ion batteries in the last decade \citep{batt_review}, battery longevity is still a major barrier for real-world applications \citep{batt_issues}. Consequently, EV drivers are more conscious about eco-driving practices \citep{evdriver_more_eco}. With the popularity of EVs such as Nissan Leaf, Chevy Bolt, and Tesla, more data on the real-world lifetimes of electric vehicles are becoming available \citep{Nissan_leaf}. Laboratory tests placed the lifetime of Nissan Leaf batteries at around 50,000 miles \citep{Idaho_nissan_leaf} while crowd-sourced data showed that Tesla batteries are capable of holding their capacity up to 400,000 miles \citep{tesla_batt}. This large difference in lifetimes may be attributed to different battery chemistries, battery architecture, as well as battery pack size. For example, NCM (Nickel Cobalt Manganese) batteries used in the Nissan Leaf and NCA (Nickel Cobalt Aluminum) batteries used by Tesla generally have a shorter lifetime than LFP (Lithium Iron Phosphate) batteries \citep{battery_chemistry}. Larger batteries would experience lower charge/discharge rates (C-rate) which reduce degradation and extend battery lifetime. These two factors will affect the real-world lifetime cost of the battery. Thus, it is also important to assess the effects of battery decay rate and battery pack size on the benefits of eco-driving with electric vehicles.

All of the above-mentioned references focused on the benefit of eco-driving in reducing energy consumption. To our knowledge, there exist only two previous studies that have addressed the role of battery degradation in optimal eco-driving strategies for pure EVs \citep{cruise_control_battery_lifetime, ev_trucks}. However, neither of these two studies considered battery degradation in the context of eco-driving at consecutive signalized intersections. They also did not address how varying the battery chemistry and battery size would affect the results. We attempt to close this gap in the literature by presenting a comprehensive study where we consider the eco-driving of a single connected electric vehicle equipped with vehicle-to-infrastructure (V2I) communication passing through a simplified control zone consisting of two signalized intersections. Our goal is to focus on minimizing both energy consumption and battery degradation using this simplified control zone as a first step without considering complicated traffic scenarios, driver behavior, or car-following models. This study is an extension of our previous work \citep{wes} which investigated the benefits of an eco-driving controller for an electric bus passing through a single signalized intersection. An eco-driving algorithm is constructed here that incorporates a battery degradation model in addition to minimizing energy consumption using dynamic programming to optimize the vehicle's speed trajectory while transiting the control zone within the specified constraints. 

There are four novel aspects to this study. First, we consider two consecutive signalized intersections at varied separations to better reflect real-world urban driving scenarios and gain a comprehensive understanding of their effect. The results obtained here can be easily extended to other practical traffic situations. Second, we incorporate various battery chemistries to include the effect of battery capacity decay rate on the benefits of eco-driving. These results provide a comprehensive understanding of the interplay between battery degradation and energy consumption on eco-driving strategies. Third, we consider two battery sizes based on the Tesla Model 3 to study their effect on the lifetime cost under eco-driving scenarios. These results can be analyzed to provide guidance on the choice of battery chemistry and battery size during the EV design process, as well as to devise an eco-driving strategy for real-world implementation. Finally, we present results from an eco-driving field test. Previous tests of eco-driving involved either retrofitting additional devices within the vehicle \citep{ead, v2c, eco_report}, or using a simulator to observe the driver's ability to follow speed advisories \citep{eco_simulator}. Here, we have conducted field tests by integrating the speed recommendations in a phone app which does not require additional devices. The app issues only audio speed advisories to the driver in order to minimize distractions. The app is convenient to use and is readily downloadable; therefore, it should facilitate broader adoption and widespread use of eco-driving.

The next section introduces the numerical model, followed by a parametric study for various signal timings and distances between the two intersections. Results are also presented for different battery capacity decay rates based on battery chemistry, as well as the effect of battery size. The cost benefits obtained by eco-driving over regular driving due to energy savings and improved battery life are analyzed and discussed.

\section{Numerical model and implementation}
The powertrain parameters are extracted from experimental testing conducted with our fuel cell buses at the Center for Fuel Cells and Batteries at the University of Delaware. Details can be found in \citep{WANG_pms} and the powertrain parameters and battery capacity were scaled down to simulate a battery-electric passenger vehicle like the Tesla Model 3. The instantaneous vehicle power demand $P$ is calculated according to: 

\begin{equation}
\begin{split}
P = ( ma + mg \sin\theta + (C_{r1} + C_{r2} v) mg \cos\theta + \\ \frac{1}{2} \rho A C_d v^2 ) \cdot v/(\eta_{trans} \eta_{motor} \eta_{invert})
\end{split}
\label{eqn:vehicle_model}
\end{equation}
where $m$ is the mass of the vehicle and $a$ is its acceleration, $g$ is the acceleration due to gravity, $\theta$ is the road inclination angle ($\theta= \arctan(grade)$), $C_{r1}$ and $C_{r2}$ are rolling resistance coefficients, $v$ is the vehicle velocity, $\rho$ is the air density, $A$ is the vehicle's frontal area, $C_d$ is the aerodynamic drag coefficient, and $\eta_{trans}$, $\eta_{motor}$ and $\eta_{invert}$ are the efficiencies of the transmission, motor and inverter, respectively. The vehicle parameters used in the current study are listed in \cref{tab:powertrain_parameters}.
\begin{table}[h]
\centering
    \begin{tabular}{|c|c|c|}
    \hline
    \textbf{Parameter} & \textbf{Symbol} & \textbf{Value}\\ \hline \hline
    Mass & $m$ & \SI{1611}{kg} \\ \hline
    Rolling resistance & $C_{r1}$ & 0.0065 \\ \hline
    Rolling resistance & $C_{r2}$ & $4.92 \times 10^{-5}$ s/m \\ \hline
    Frontal area & $A$ & \SI{2.22}{m^2} \\ \hline
    Air drag coefficient & $C_d$ & 0.23\\ \hline
    Battery capacity & $C_{batt}$ & 54 kWh \\ \hline
    Transmission efficiency & $\eta_{trans}$ & 0.8536 \\ \hline
    Inverter efficiency & $\eta_{invert}$ & 0.95 \\ \hline
    \end{tabular}
    \caption{Parameters used in this study to simulate a passenger electric vehicle.}
    \label{tab:powertrain_parameters}
\end{table}

The battery degradation model employed corresponds to the LFP chemistry developed in \citep{LiFePO4} which has a longer lifetime compared to NCM and NCA batteries used extensively in the current generation of electric vehicles \citep{battery_chemistry}. We will show subsequently that varying the battery decay rate does not greatly affect the optimal speed trajectories; however, the battery lifetime cost is a function of the type of battery used. The battery degradation model is based on the total discharged Ah throughput over the lifetime of the battery:

\begin{equation}
	A_{total}(c, T) = \left[ \frac{\Delta Q_b}{M(c) exp\left( \frac{-E_a(c)}{RT} \right)} \right]^{\frac{1}{z}}
	\label{eqn:batt_ah}
\end{equation}
where $A_{total}$ is the discharged Ah throughput depending on C-rate, $\Delta Q_b$ is the percentage capacity loss which is set to 20\% to indicate end-of-life, $c$ is the C-rate, $R$ is the ideal gas constant ($R$ = 8.3144 $J/K/mol$), and $T$ is temperature ($T$ = 298.15 K). $M(c)$ is the pre-exponential factor which is a function of the C-rate.  The activation energy $E_a$ and the power-law factor $z$ are fitting parameters ($z$ = 0.55). The battery state-of-health (SOH) decay rate $\overset{\boldsymbol{.}}{B}_{soh}$ is calculated as:
\begin{equation}\label{eq:batt_soh}
	\overset{\boldsymbol{.}}{B}_{soh}(t) = - \frac{|I(t)|}{2A_{total}(c, T)}
\end{equation}
where $I(t)$ is the current. A detailed derivation can be found in \citep{batt_decay_model}. It should be noted that there are many factors that could affect battery degradation under real-world driving scenarios and the empirical model employed here is not meant to be exclusive. The methodology presented here can be readily implemented with alternative degradation models as well. Nevertheless, the specific model employed here is representative of real-world battery degradation and the conclusions presented in this study should hold in general.

The effect of driving behavior on energy consumption and battery degradation while passing through a control zone consisting of two consecutive signalized intersections is investigated in this study. This is a generalized approach that can be extended to multiple intersections by sequentially considering two consecutive intersections at a time. The control zone extends to 100 m before the first light and 100 m after the second light such that the eco-driving controller starts to receive signal-timing information as it approaches within 100 m of the first light and reaches the speed limit 100 m after the second light. The speed limit is set to 88.5 km/h (55 mph) and the red and green signal durations are each set to 30 seconds. The distance between the two traffic lights is varied from 200 m to 800 m and different signal timing combinations are studied to evaluate their effect on the optimal speed trajectories. Dynamic programming (DP) was used to optimize the velocity trajectory of the eco-driving vehicle as it transits the control zone with energy consumption and battery degradation as the cost objectives. The cost benefits resulting from eco-driving are compared with a regular driver who is assumed to decelerate at a constant rate if they see a red light within 75 meters of the intersection, and accelerate linearly to 88.5 km/h (55 mph) once it turns green. The deceleration rate was calculated based on the vehicle's distance to the light and its current speed such that the vehicle will come to a complete stop at the light. The maximum allowed acceleration and deceleration rates are set to  2 m/s$^2$ \citep{driver_comfort_accl} and -4 m/s$^2$ \citep{driver_comfort_decel}, respectively.

The state space for DP includes speed and time which provides the benefit that the control zone exit times do not need to be known \emph{a priori} \citep{uncertain_traffic_signal_timing}. For a given distance between the two lights and their signal timing, the speed trajectory of the regular driver was obtained first and the trip time was then used as the upper limit for the eco-driving controller. The eco-controller's DP cost function includes the energy consumption (cost of electricity) and the battery decay (battery replacement cost): 

\begin{equation}
min\quad J = \int_{0}^{T} (c_{e}P_{batt}  + c_{bat} \overset{\boldsymbol{.}}{B}_{soh})\ dt \\ 
\label{eqn:objective_func}
\end{equation}
where $c_{e}$ is the electricity cost of $\$0.12/kWh$, $P_{batt}$ is the battery gross power, $c_{bat}$ is the capital cost of the battery pack based on the DOE 2022 target price of $\$125/kWh$, and $\overset{\boldsymbol{.}}{B}_{soh}$ is the battery SOH decay rate. The battery size is set to 54 $kWh$ based on the 2019 version of the Tesla Model 3 with a standard-range battery pack.

Numerical issues were encountered when the parameter combination was such that the regular driver needed to apply the brake only for a brief period within the control zone which maintained the average speed close to the speed limit. This poses a numerical issue for the eco-driving algorithm because the DP has to find a solution path at the very edge of the feasible state space. This would require a much smaller time step for the simulation to execute successfully which adds to the overall computational time. An alternative solution is to add a small buffer to the travel time; the results showed that varying the travel time by $\pm 3$\% of the regular driver has a minor effect on the cost. Another solution is to increase the time resolution along the speed limit boundary in state space by decreasing the time step for speeds that are close to the speed limit. This reduced the numerical error and helped DP to find a path close to the boundary. Adjusting the resolution ratio between time and speed also helped which was used in this study. This is the easiest to implement but it does increase computational time which could make it less practical for real-time applications, although it is acceptable for offline simulations.  Intuitively, the time resolution should be set to equal the distance step divided by the speed limit, which would be the shortest time needed to explore all the feasible speeds. However, this did not reduce the numerical error for all cases. Therefore, the timestep was set to half or smaller for most simulations to successfully complete.

\section{App development and field test}
\begin{figure}
	\centering
	\includegraphics[width=0.9\columnwidth]{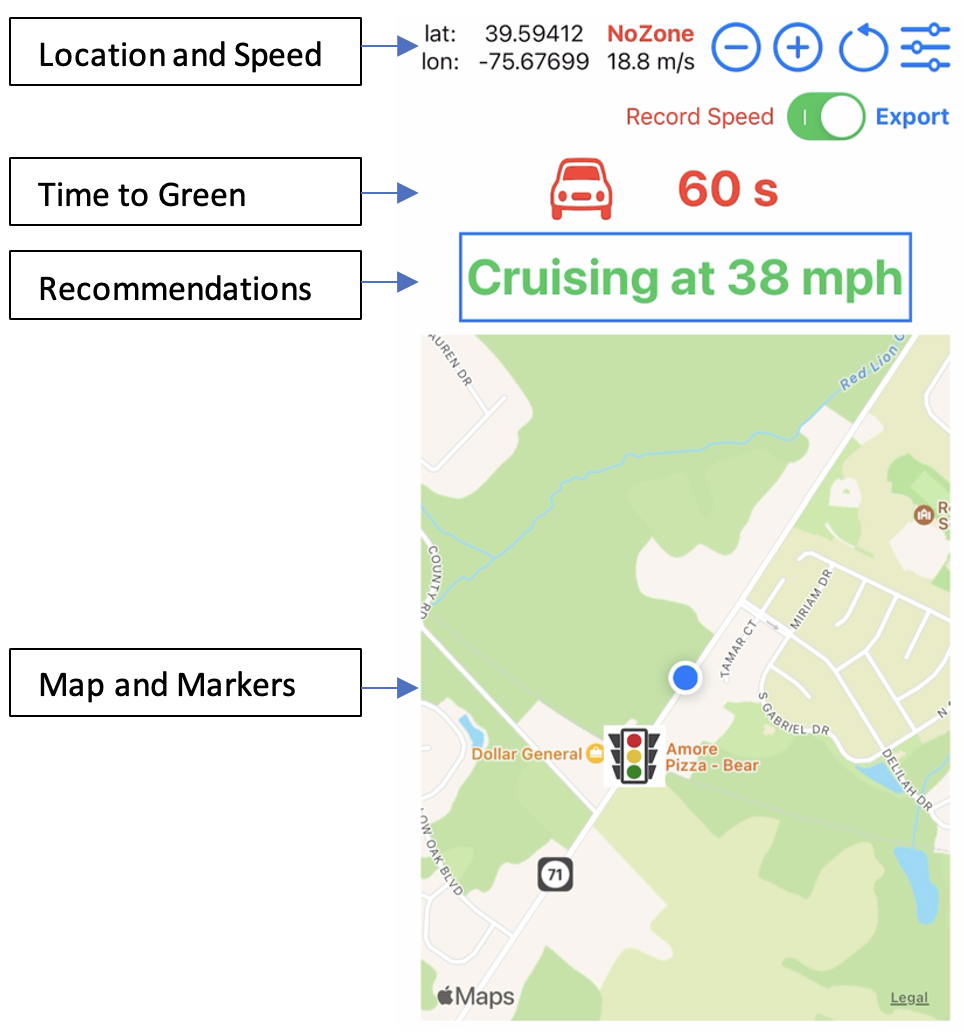}
	\caption{The iPhone app interface developed to test eco-driving for connected electric vehicles. GPS locations and vehicle speed are obtained using the phone's GPS data. Light timing, recommended actions, and a map are also shown. Audio alerts are issued to provide speed recommendations to the driver.}
	\label{fig:iTAP_app}
\end{figure}

A phone app was developed and implemented in a field test to validate the results from simulations. The app as shown in \cref{fig:iTAP_app} was developed using SwiftUI and was installed on an iPhone. The vehicle location and speed are updated using GPS data from the phone at 1 Hz. The time to green for the next traffic light is shown in the middle with the recommended actions for the driver. Light phasing information is readily available from commercial vendors with high accuracy and low latency. In our field test, we have assumed that such phasing information can be retrieved by the phone and used to calculate speed recommendations. A map with markers for the vehicle (blue dot) and approaching traffic lights (traffic light icon) is shown at the bottom. The eco-driving results presented here are obtained for a simplified rule-based strategy for real-world implementation that advises the driver to brake or accelerate to a certain cruising speed according to the distance to and timing of the upcoming traffic light. The app then issues voice recommendations as the vehicle approaches the signalized intersection so as to minimize distractions to the driver. The timing of two upcoming traffic signals are received and analyzed to optimize the speed profile to pass through them. The vehicle's speed is then recorded and exported for analysis. The intersections and lights were selected at a location that had good visual clearance and minimized disturbances to the normal flow of traffic. The field tests were carried out around dawn using a regular gasoline sedan (2014 Honda Accord EX) when the traffic volume is low.

\section{Results and Discussion}
This section presents results in which the role of two parameters, distance between the two signals and signal timing combination, is explored. The speed trajectories between eco-driving and regular driving are compared along with the resulting energy and battery decay costs.

\subsection{Effect of eco-driving speed trajectories on cost}
\begin{figure}
	\centering
	\includegraphics[width=0.9\columnwidth]{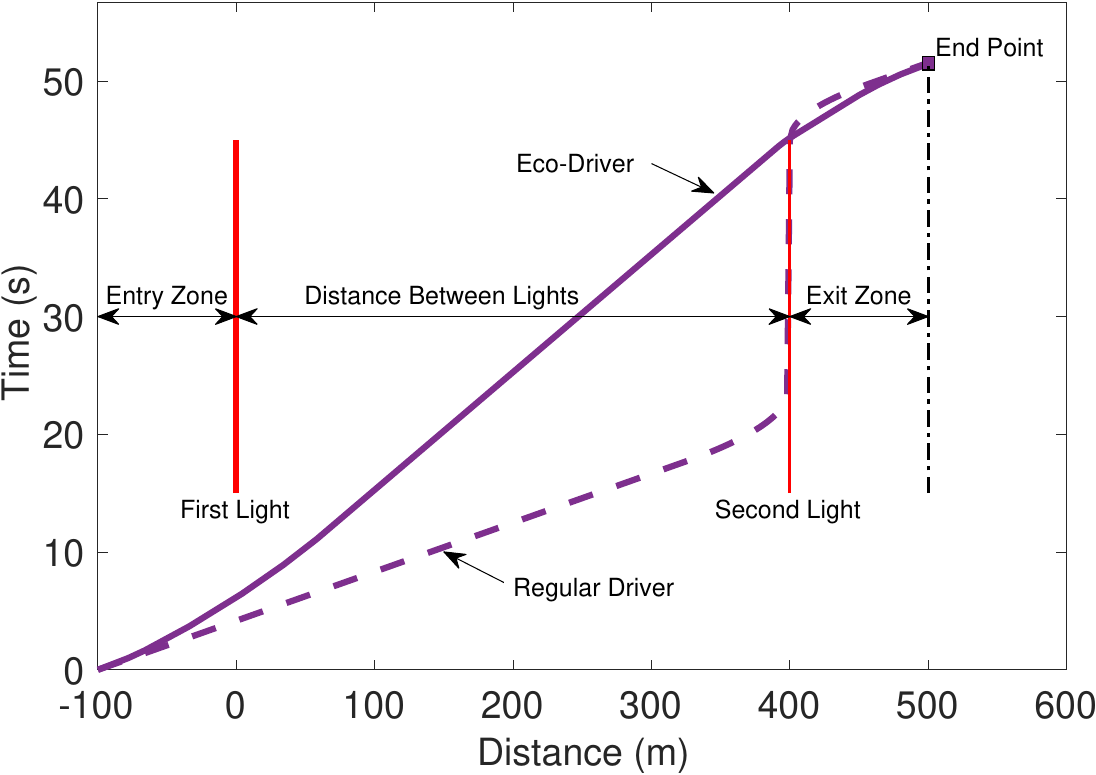}
	\caption{An illustrative example of a vehicle trajectory through time and distance. The control zone commences 100 m upstream of the first light and the ends 100 m downstream of the second light. In this example, the distance between the two lights is 400 m, and they have the same signal timing. Both the distance and cycle timing will be varied to evaluate their effect on the benefits of eco-driving. The solid and dashed lines show the trajectories of the eco-driver and regular driver, respectively.}
	\label{fig:15_15_trajectory_illustration}
\end{figure}

The signal timing of the two lights is defined by $[x,y]$ where $x$ and $y$ refer to the time to red in seconds for the first and second light, respectively, after the vehicle enters the 100 m control zone. For example, [0 15] indicates that the first light turns red as soon as the vehicle enters the control zone, and the second light turns red 15 seconds after the vehicle enters the control zone. A sample vehicle trajectory in time-distance space is shown in \cref{fig:15_15_trajectory_illustration} to explain all the elements in the subsequent trajectory plots.

\begin{figure*}
	\centering
	\includegraphics[width=0.9\textwidth]{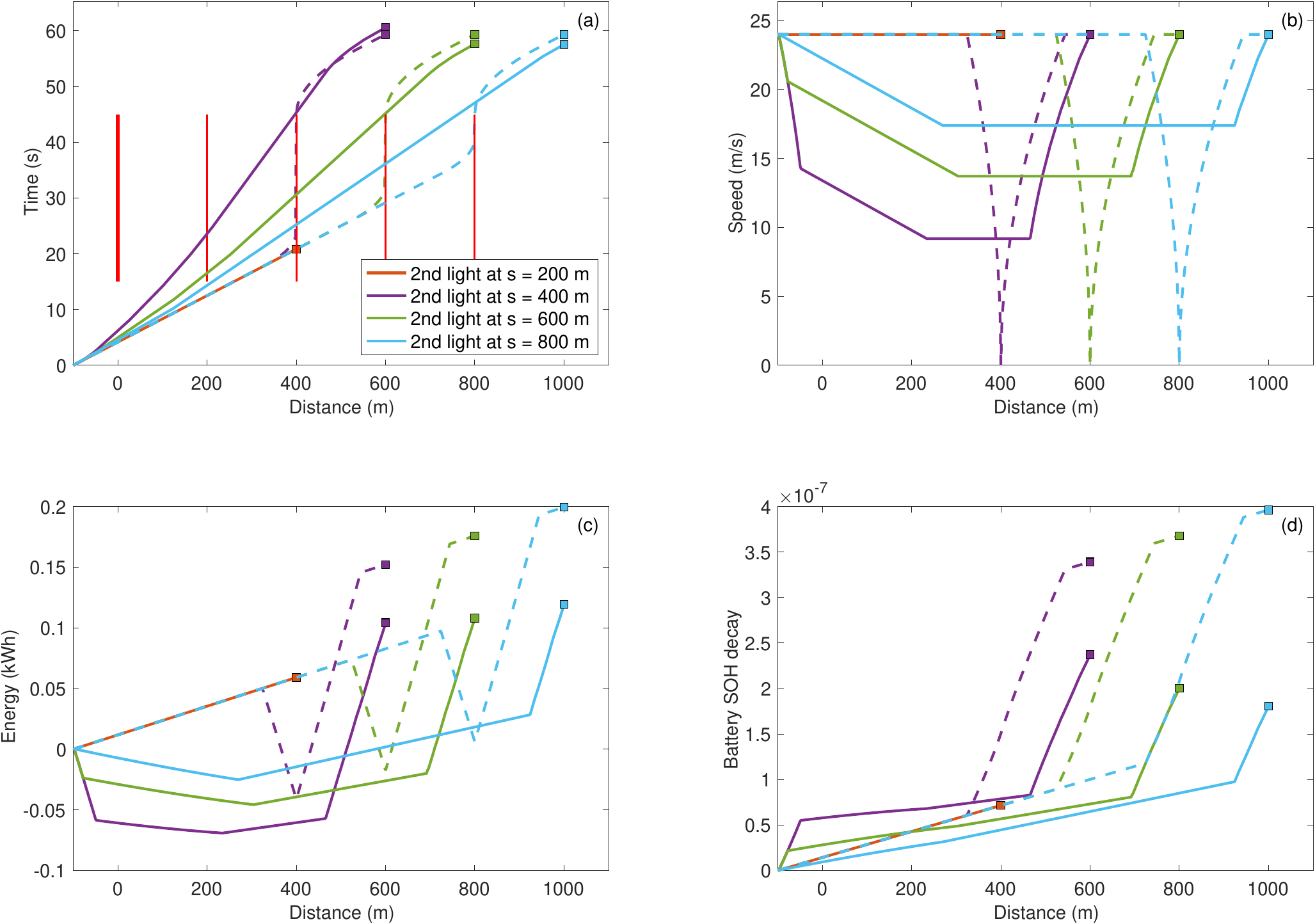}
	\caption{Results comparing eco-driving with regular driving for signal combination [15 15] when both lights turn red 15 seconds after the vehicle enters the control zone. \full \ Eco-driver; \dashed \ Regular driver. (a) Trajectories through time and distance; (b) speed trajectories; (c) energy consumption; and (d) battery capacity decay.}
	\label{fig:15_15_trajectory}
\end{figure*}

\begin{figure*}
	\centering
	\includegraphics[width=0.9\textwidth]{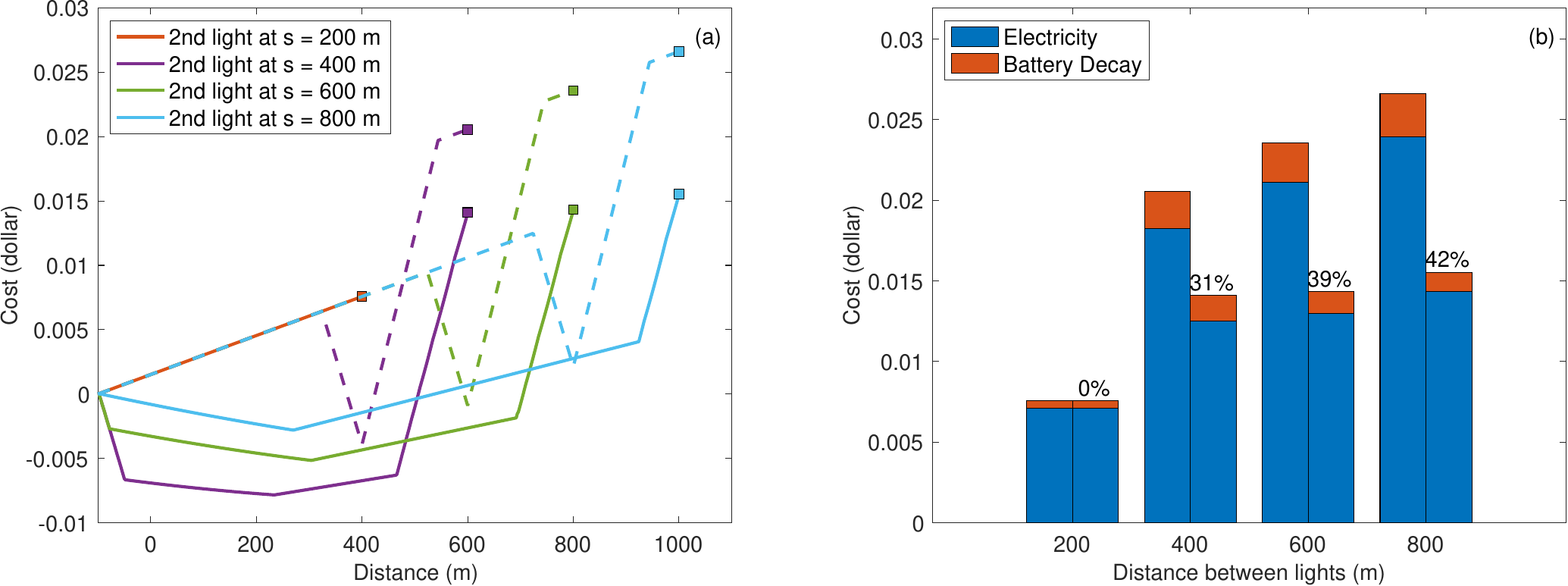}
	\caption{Results comparing eco-driving with regular driving for signal combination [15 15]. (a) Total cost,  \full \ Eco-driver, \dashed \ Regular driver; and (b) electricity and battery cost comparison with the overall cost savings denoted as a \%.}
	\label{fig:15_15_cost}
\end{figure*}

The results for the signal timing combination of [15 15] are shown in \cref{fig:15_15_trajectory}. The first vertical bar corresponds to the location of the first light 100 m into the control zone. This light turns red 15 seconds after the vehicle enters the control zone, and remains red for another 30 seconds. The four additional vertical bars correspond to the location of the second light situated at $s=$ 200, 400, 600 and 800 m, respectively, after the first light. Each second light also turns red 15 seconds after the vehicle enters the control zone for this signal combination and remains red for another 30 seconds. There is a 200 m exit zone after the second light. The trajectory of the regular driver is denoted by the dashed line, and that of the eco-driver by the solid line. The trajectories corresponding to each $s$ value are indicated by different colors, and the end of each trajectory is denoted by a square dot whose color is matched to its corresponding trajectory. \cref{fig:15_15_trajectory}a shows that the first light does not affect the trajectory of the regular driver for any value of $s$. It is also seen that the eco-driver follows the same trajectory as the regular driver for $s=200$ m. However, the regular driver must brake to a complete stop for $s=$ 400, 600 and 800 m as shown in \cref{fig:15_15_trajectory}b. In contrast, since the eco-controller knows the timing of the second light \emph{a priori}, the algorithm optimizes the overall cost to reduce energy consumption and battery degradation by recommending a speed trajectory that decelerates the vehicle immediately upon entering the control zone, and then makes it cruise at a lower speed such that it passes the second light exactly when it turns green. This dramatically reduces the energy consumption (\cref{fig:15_15_trajectory}c) and battery decay (\cref{fig:15_15_trajectory}d), and the cost for all three cases ($s=$ 400, 600 and 800 m) as shown in \cref{fig:15_15_cost}. For $s=800$ m, the energy saving is around 40\% while the battery decay reduction is around 54\% for an overall cost reduction of 42\%. 

\begin{figure*}
	\centering
	\includegraphics[width=0.9\textwidth]{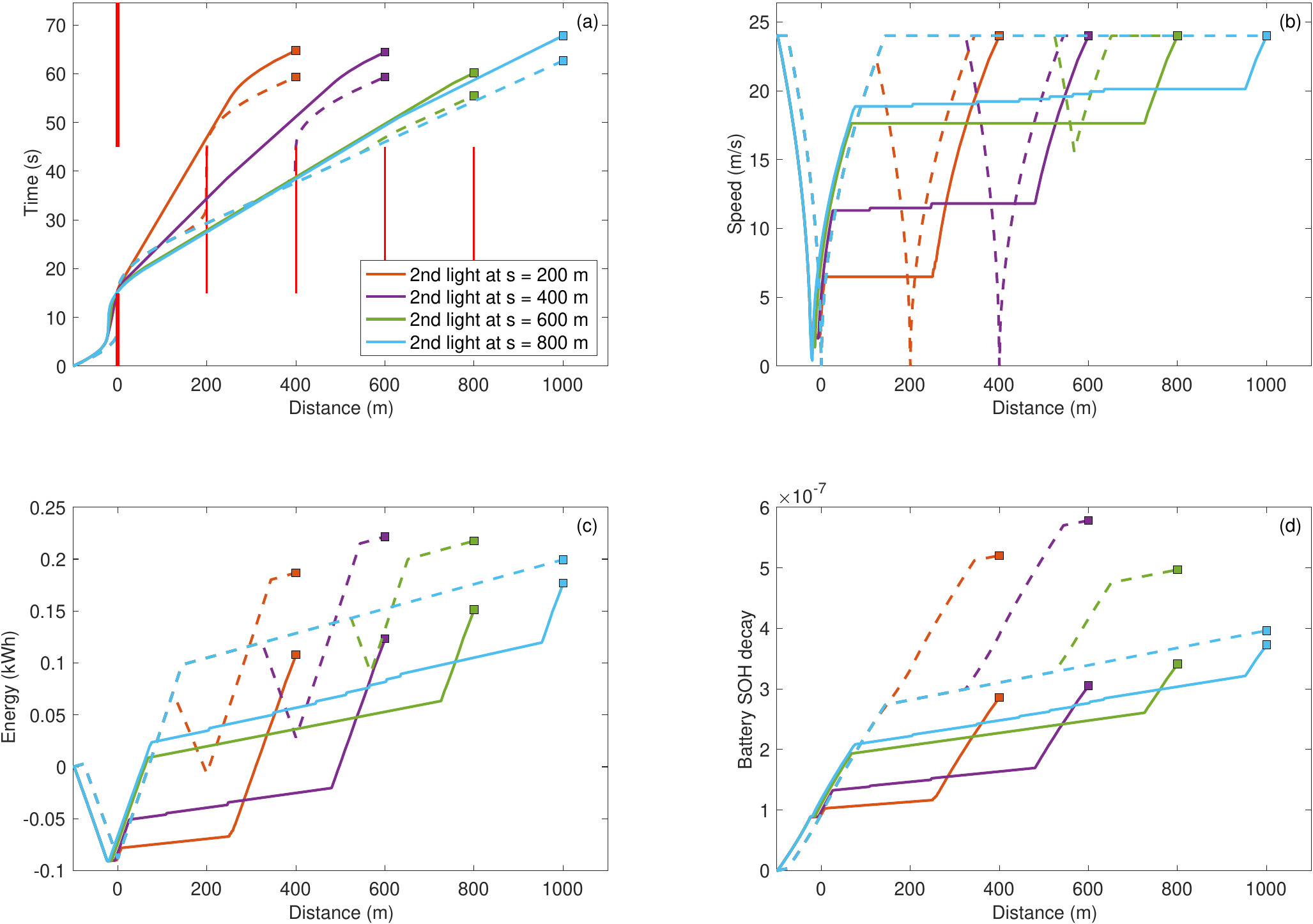}
	\caption{Results comparing eco-driving with regular driving for signal combination [-15 15] when the first and second lights turn red 15 seconds before and after the vehicle enters the control zone, respectively. \full \ Eco-driver; \dashed \ Regular driver. (a) Trajectories through time and distance; (b) speed trajectories; (c) energy consumption; and (d) battery capacity decay.}
	\label{fig:-15_15_trajectory}
\end{figure*}

\begin{figure*}
	\centering
	\includegraphics[width=0.9\textwidth]{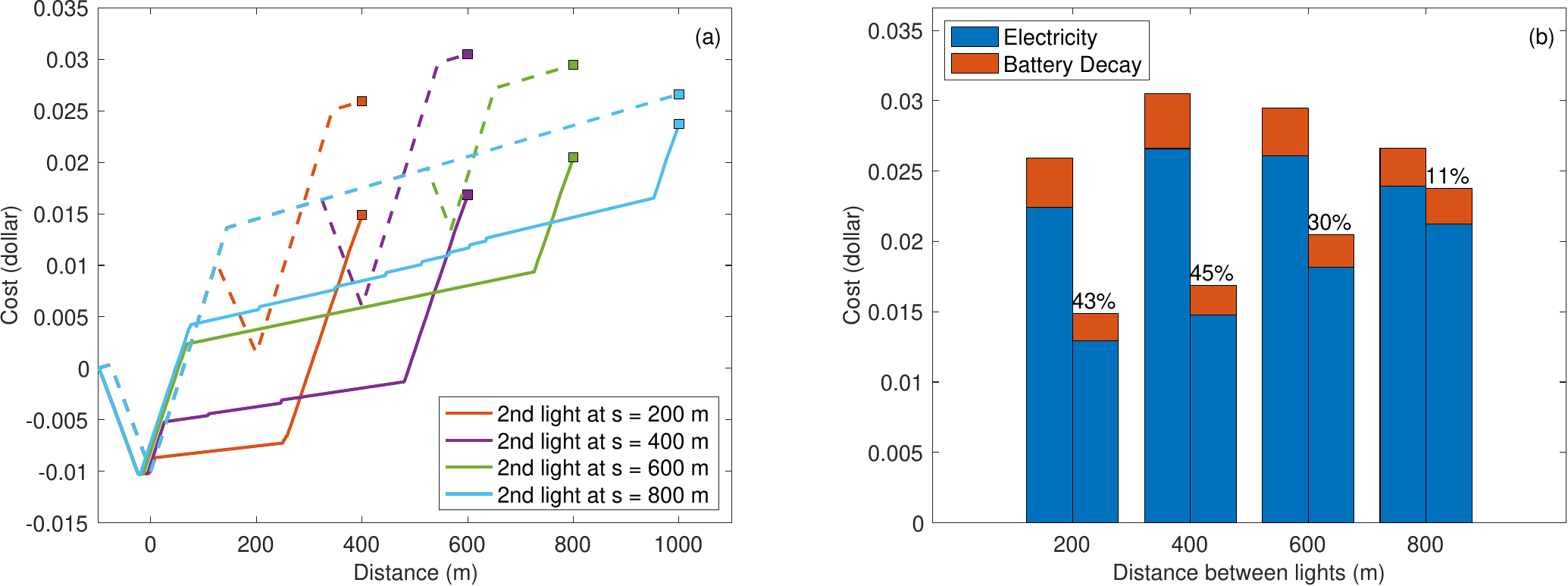}
	\caption{Results comparing eco-driving with regular driving for signal combination [-15 15]. (a) Total cost,  \full \ Eco-driver, \dashed \ Regular driver; and (b) electricity and battery cost comparison with the overall cost savings denoted as a \%.}
	\label{fig:-15_15_cost}
\end{figure*}

Results for a second signal timing combination corresponding to [-15 15] are shown in \cref{fig:-15_15_trajectory}. In this case, the regular driver must brake to a complete stop for the first light for all values of $s$. The eco-driver is also forced to decelerate; however, the eco-driver handles this scenario more intelligently by braking earlier and more gradually which eliminates the need for a complete stop. After the first light turns green, the regular driver immediately accelerates to the speed limit and is then forced to brake again to a complete stop for the second red light. On the other hand, the eco-controller provides a speed trajectory that allows the vehicle to cruise through the first light just as it turns green. Since the timing is known \emph{a priori}, the eco-controller only accelerates to a lower speed which enables it to cruise through the second intersection as well before finally accelerating to the speed limit at the exit. \cref{fig:-15_15_cost} shows that the reduction in total cost is about 30-45\% for $s=$ 200, 400 and 600 m, while it is around 11\% for $s=800$ m. The eco-driver also experiences a shorter total travel time which is an extra benefit. 

The greatest reductions in cost are due to the avoidance of hard braking and acceleration events which are associated with higher battery C-rates as well as lower efficiency for the powertrain components such as the traction motor and the transmission. Similar speed trajectories and cost reductions were observed for other signal timing combinations. In general, the optimal speed trajectories consistently favor an earlier and more gradual deceleration which prevents braking to a complete stop at either light, and permits cruising at lower speeds through the control zone which minimizes aerodynamic power loss.

A comprehensive set of 16 signal-time combinations was simulated and the corresponding cost reductions are shown in \cref{table:cost_table}. The cost reduction ranges from 0\% to 47\% depending on the signal timing. An average reduction of around 23\% was observed for different timing combinations for all values of $s$. These results confirm the effectiveness of eco-driving under different scenarios.

\begin{table}
\caption{Cost reduction (in \%) from eco-driving for 16 signal-timing combinations and the distance between lights $s=$ 200, 400, 600 and 800 $m$. The signal timing combination between the two lights is defined by $[x,y]$ where $x$ and $y$ refer to the time to red for the first and second light, respectively, after the vehicle enters the control zone. An average cost reduction of around 23\% was observed.}
\centering
\begin{tabular}{|rr|r|r|r|r|}
\hline
\multicolumn{2}{|c|}{\textbf{Timing}}&\textbf{200 m}&\textbf{400 m}&\textbf{600 m}&\textbf{800 m}\\\hline
\textbf{-30} & \textbf{-30}&0.0&0.0&0.0&40.0\\\hline
\textbf{-30} & \textbf{-15}&47.2&0.0&0.0&0.0\\\hline
\textbf{-30} & \textbf{0}&23.5&38.8&40.6&0.0\\\hline
\textbf{-30} & \textbf{15}&0.0&0.0&0.0&31.4\\\hline
\textbf{-15} & \textbf{-30}&9.8&46.6&45.2&41.9\\\hline
\textbf{-15} & \textbf{-15}&9.8&11.2&47.7&45.6\\\hline
\textbf{-15} & \textbf{0}&36.6&11.2&11.9&10.8\\\hline
\textbf{-15} & \textbf{15}&42.6&44.7&30.5&10.8\\\hline
\textbf{0} & \textbf{-30}&42.3&45.0&31.1&13.0\\\hline
\textbf{0} & \textbf{-15}&9.6&46.5&45.1&41.8\\\hline
\textbf{0} & \textbf{0}&10.1&11.7&47.6&45.3\\\hline
\textbf{0} & \textbf{15}&36.8&11.7&13.2&12.2\\\hline
\textbf{15} & \textbf{-30}&0.0&0.0&0.0&40.0\\\hline
\textbf{15} & \textbf{-15}&47.2&0.0&0.0&0.0\\\hline
\textbf{15} & \textbf{0}&23.5&38.8&40.6&0.0\\\hline
\textbf{15} & \textbf{15}&0.0&31.4&39.2&41.6\\\hline
\multicolumn{2}{|c|}{\textbf{Average}}&21.2&21.1&24.5&23.4\\\hline
\end{tabular}
\label{table:cost_table}           
\end{table} 

\subsection{Effect of battery capacity decay rate}
\begin{figure}
	\centering
	\includegraphics[width=0.9\columnwidth]{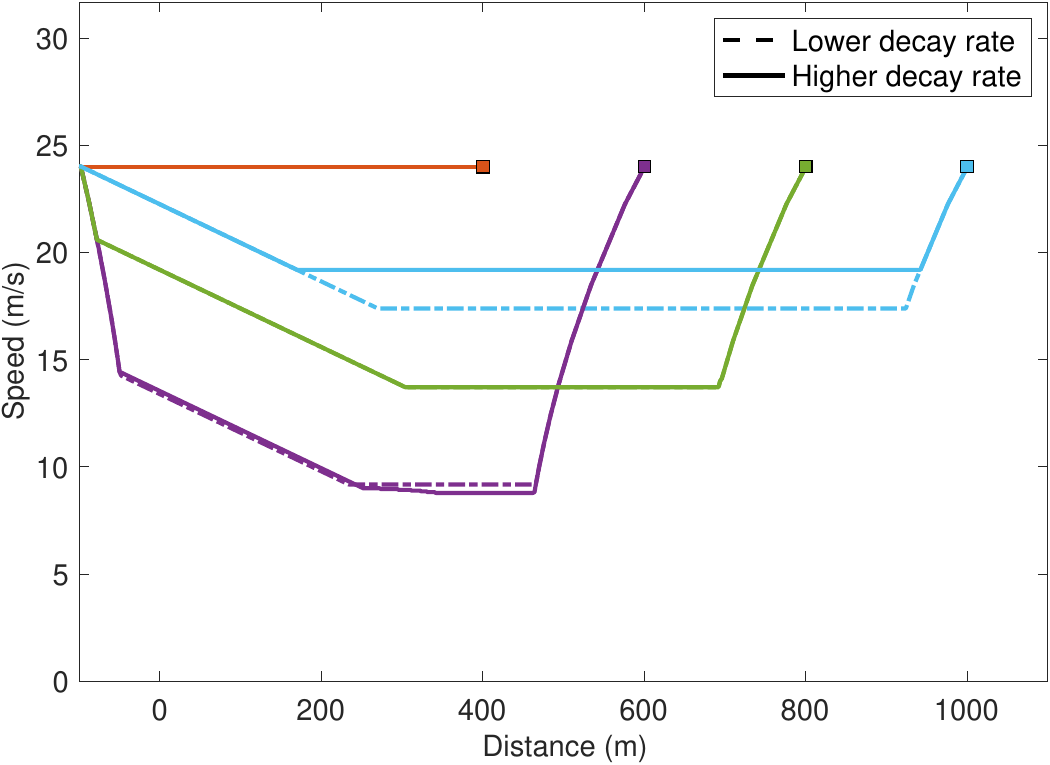}
	\caption{Eco-driving speed trajectories for the [15 15] case for two battery decay rates. The dashed lines refer to the baseline case shown previously in \cref{fig:15_15_cost}b, and the solid lines refer to the case with 10x greater battery decay rate.}
	\label{fig:15_15_batt_decay_rates_trajectories}
\end{figure}

Since the battery capacity decay rate depends heavily on battery chemistry and architecture, cases with much higher battery decay rates were also investigated. \cref{fig:15_15_batt_decay_rates_trajectories} shows the eco-driving speed trajectories for the [15 15] case for two battery decay rates: the baseline case shown earlier in \cref{fig:15_15_trajectory}b (solid lines), and a corresponding set of graphs with the battery decay rate magnified by a factor of 10 (dashed lines). The factor of 10 was selected based on \citep{battery_chemistry} which showed that the NCA battery decays 10 times faster than the LFP battery. Surprisingly, the optimal speed trajectories for eco-driving are found to be quite insensitive to the battery decay rate as shown in \cref{fig:15_15_batt_decay_rates_trajectories}. Thus, eco-driving is seen to yield similar speed profiles regardless of battery chemistry. Since the optimal speed profiles are so similar despite increasing the battery capacity decay so drastically, it indicates that better driving behaviors like gradual braking that lead to greater energy savings are more important in reducing overall costs.

\begin{figure*}
	\centering
	\includegraphics[width=0.9\textwidth]{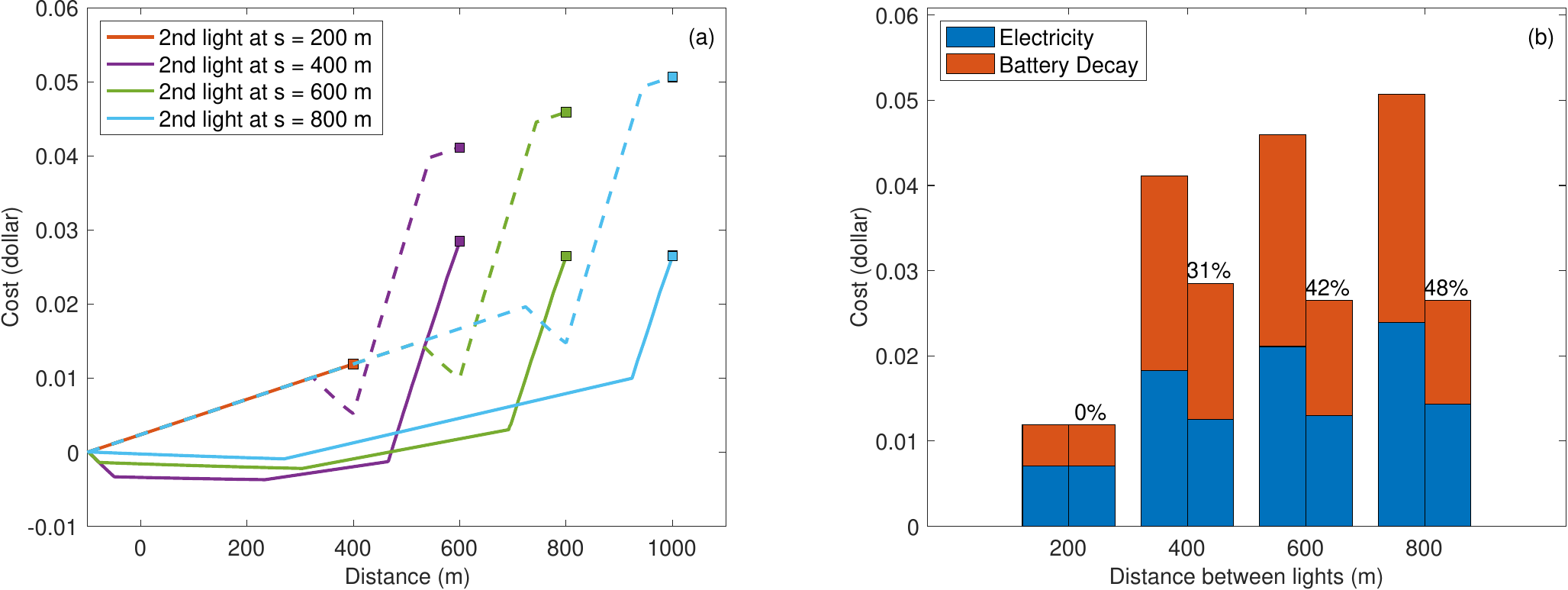}
	\caption{Results comparing eco-driving with regular driving for signal combination [15 15] for the 10x battery decay rate case. (a) Total cost,  \full \ Eco-driver, \dashed \ Regular driver; and (b) electricity and battery cost comparison with the overall cost savings denoted as a \%.}
	\label{fig:15_15_batt_decay_rates_cost}
\end{figure*}

Increasing the decay rate by 10x does make the battery decay cost higher compared to energy consumption as shown in \cref{fig:15_15_batt_decay_rates_cost}. The factor of ten is from \citep{battery_chemistry} which showed that NCA batteries decays approximately ten times faster than LFP batteries. In this case, the costs are almost equal between energy and battery capacity decay. The total cost reduction is up to 50\% for the 800 $m$ distance between the intersections due to the increased proportion of battery decay in the total cost.

\subsection{Effect of battery size}
\begin{figure}
	\centering
	\includegraphics[width=0.9\columnwidth]{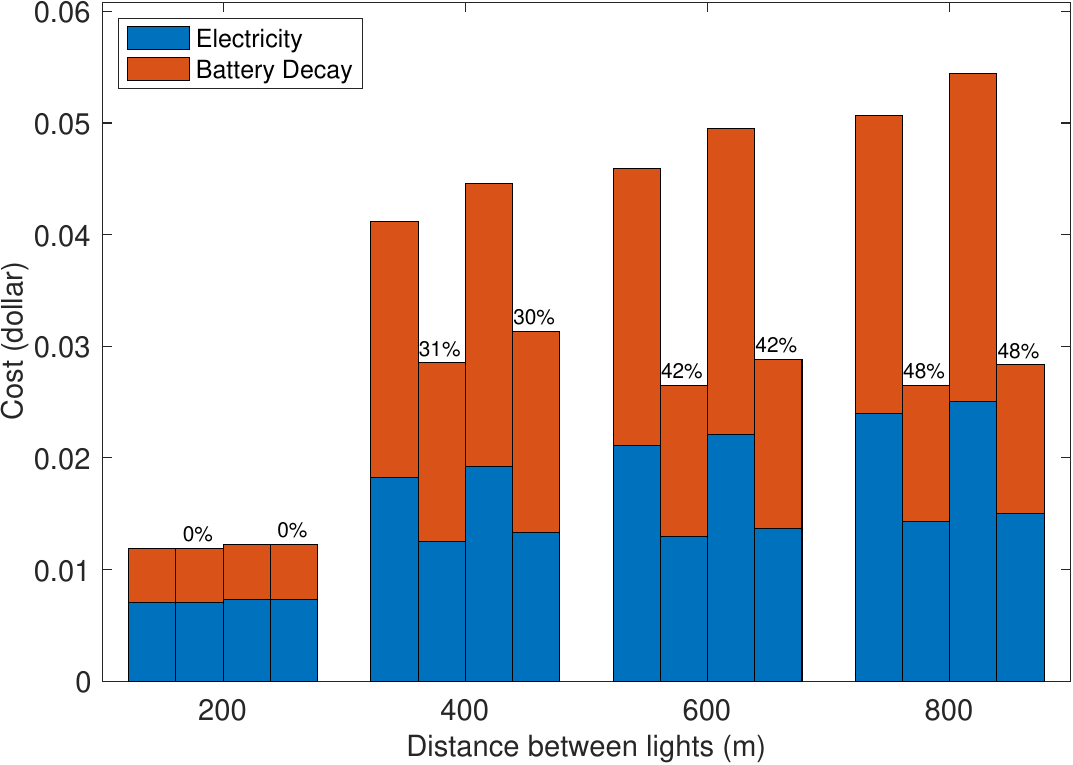}
	\caption{Cost comparisons for eco-driving with regular driving for signal combination [15 15] for battery sizes of 54 and 75 $kWh$. The columns in each group from left to right are: regular/54 $kWh$; eco/54 $kWh$; regular/75 $kWh$; and eco/75 $kWh$.}
	\label{fig:15_15_cost_different_battery_sizes}
\end{figure}

Battery size will affect the vehicle weight. But more importantly, it has a direct impact on the battery's charge/discharge rates which will in turn affect its capacity decay rate. Thus, it is useful to explore how battery size will influence eco-driving behavior. The baseline battery size and the corresponding vehicle weight in this study are based off the Tesla Model 3. Since Tesla vehicles use the NCA battery, the higher decay rate mentioned in the previous section was employed in these simulations. The 2020 standard-range version has a 54 $kWh$ battery and weighs 1611 $kg$ while the long-range version has a 75 $kWh$ battery and weighs 1726 $kg$. The resulting cost comparisons are shown in \cref{fig:15_15_cost_different_battery_sizes}. For both the regular and eco-driving cases, the larger battery corresponds to a greater vehicle weight which leads to higher energy consumption as shown by the slight increase in electricity cost indicated by the blue bars. The larger battery should also have a lower capacity decay due to lower C-rates, but the higher capital cost of the larger battery negates this benefit as indicated by the orange bars. The total cost reduction due to eco-driving is similar for either battery size and ranges from 30-45\% for the three higher $s$ values.

\begin{figure}
	\centering
	\includegraphics[width=0.9\columnwidth]{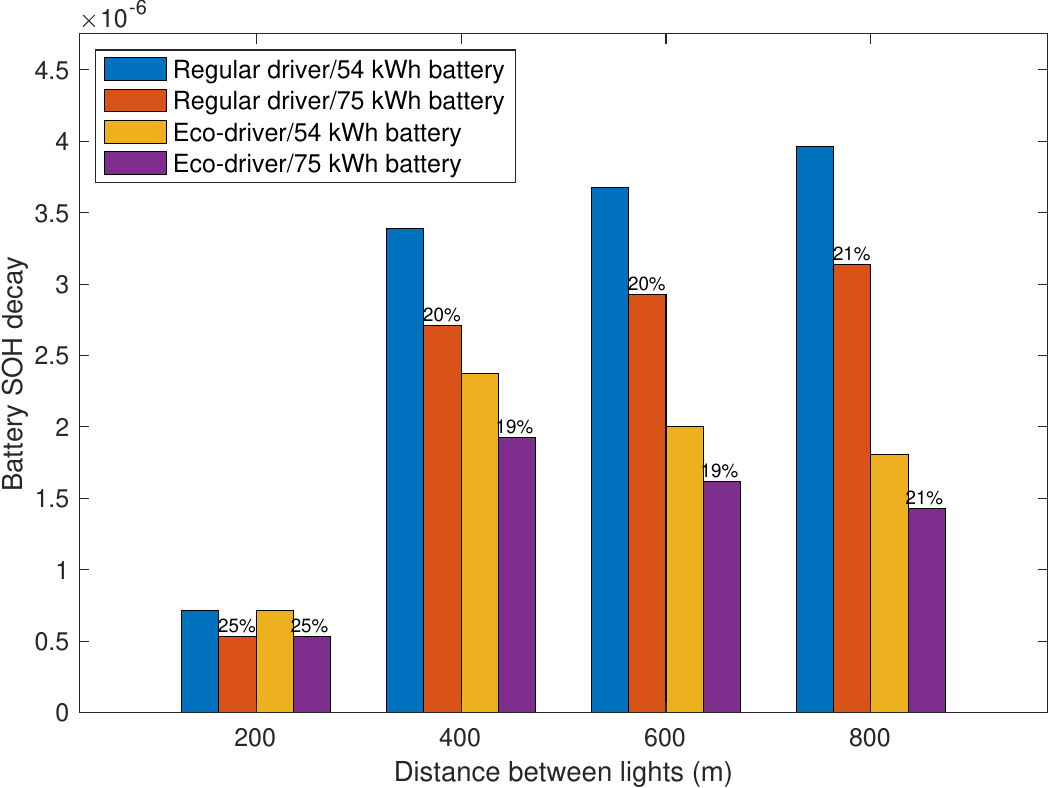}
	\caption{Comparison of battery capacity decay for battery sizes of 54 and 75 $kWh$ for signal combination [15 15]. The columns in each group from left to right are: regular/54 $kWh$; regular/75 $kWh$; eco/54 $kWh$; and eco/75 $kWh$. The percentage reductions indicate the differences between the smaller and larger battery packs.}
	\label{fig:15_15_decay_different_battery_sizes}
\end{figure}

The energy and battery decay cost differences due to vehicle weight are insignificant because the additional battery weight is relatively small compared to the curb weight of the vehicle. However, the battery size directly affects the C-rate and thus the capacity decay. The capacity decay for the two battery sizes for the signal combination of [15 15] is shown in \cref{fig:15_15_decay_different_battery_sizes} and results for all other signal-timing combinations are summarized in \cref{table:decay_table_different_battery_sizes}. The reduction in battery capacity decay for the larger battery is more than 20\% compared to the smaller battery. This translates to a 25\% increase in battery lifetime which can be exploited to support grid-integrated applications like V2G.

\begin{table}
\caption{Reduction of battery capacity decay (in \%) when battery size is increased from 54 to 75 $kWh$ for 16 signal-timing combinations and four values of distance between lights $s$. For each $s$, the left and right columns indicate the decay reduction for regular and eco-driving scenarios, respectively. An average decay reduction of around 21\% was observed for the vehicle with the larger battery.}
\centering
\resizebox{0.5\textwidth}{!}{\begin{tabular}{|r@{}r|c|c|c|c|c|c|c|c|}
\hline
\multicolumn{2}{|c|}{\textbf{Timing}}&\multicolumn{2}{c|}{\textbf{200 m}}&\multicolumn{2}{c|}{\textbf{400 m}}&\multicolumn{2}{c|}{\textbf{600 m}}&\multicolumn{2}{c|}{\textbf{800 m}} \\\hline
\textbf{-30} & \textbf{-30}&25.5&25.5&25.5&25.5&25.5&25.5&20.8&19.6\\\hline
\textbf{-30} & \textbf{-15}&19.5&19.7&25.5&25.5&25.5&25.5&25.5&25.5\\\hline
\textbf{-30} & \textbf{0}&19.5&18.6&20.0&19.2&20.4&21.7&25.5&25.5\\\hline
\textbf{-30} & \textbf{15}&25.5&25.5&25.5&25.5&25.5&25.5&20.0&18.8\\\hline
\textbf{-15} & \textbf{-30}&19.5&19.2&19.1&19.2&19.4&19.6&19.7&20.1\\\hline
\textbf{-15} & \textbf{-15}&19.5&19.2&20.0&19.7&20.9&19.4&19.7&19.8\\\hline
\textbf{-15} & \textbf{0}&18.8&19.2&20.0&19.7&20.5&20.1&20.8&20.4\\\hline
\textbf{-15} & \textbf{15}&19.0&19.0&19.1&19.4&19.4&20.0&20.8&20.4\\\hline
\textbf{0} & \textbf{-30}&19.0&19.0&19.1&19.3&19.4&20.0&20.8&20.4\\\hline
\textbf{0} & \textbf{-15}&19.5&19.3&19.1&19.2&19.4&19.6&19.7&20.1\\\hline
\textbf{0} & \textbf{0}&19.5&19.2&20.0&19.7&20.9&19.4&19.7&19.8\\\hline
\textbf{0} & \textbf{15}&18.8&19.1&20.0&19.7&20.5&20.1&20.8&20.4\\\hline
\textbf{15} & \textbf{-30}&25.5&25.5&25.5&25.5&25.5&25.5&20.8&19.6\\\hline
\textbf{15} & \textbf{-15}&19.5&19.7&25.5&25.5&25.5&25.5&25.5&25.5\\\hline
\textbf{15} & \textbf{0}&19.5&18.6&20.0&19.2&20.4&21.7&25.5&25.5\\\hline
\textbf{15} & \textbf{15}&25.5&25.5&20.0&18.8&20.5&19.4&20.8&21.0\\\hline
\multicolumn{2}{|c|}{\textbf{Average}}&20.9&20.7&21.5&21.3&21.8&21.8&21.6&21.4\\\hline
\end{tabular}}
\label{table:decay_table_different_battery_sizes}    
\end{table}

\subsection{Field test results}
\begin{figure}
	\centering
	\includegraphics[width=0.9\columnwidth]{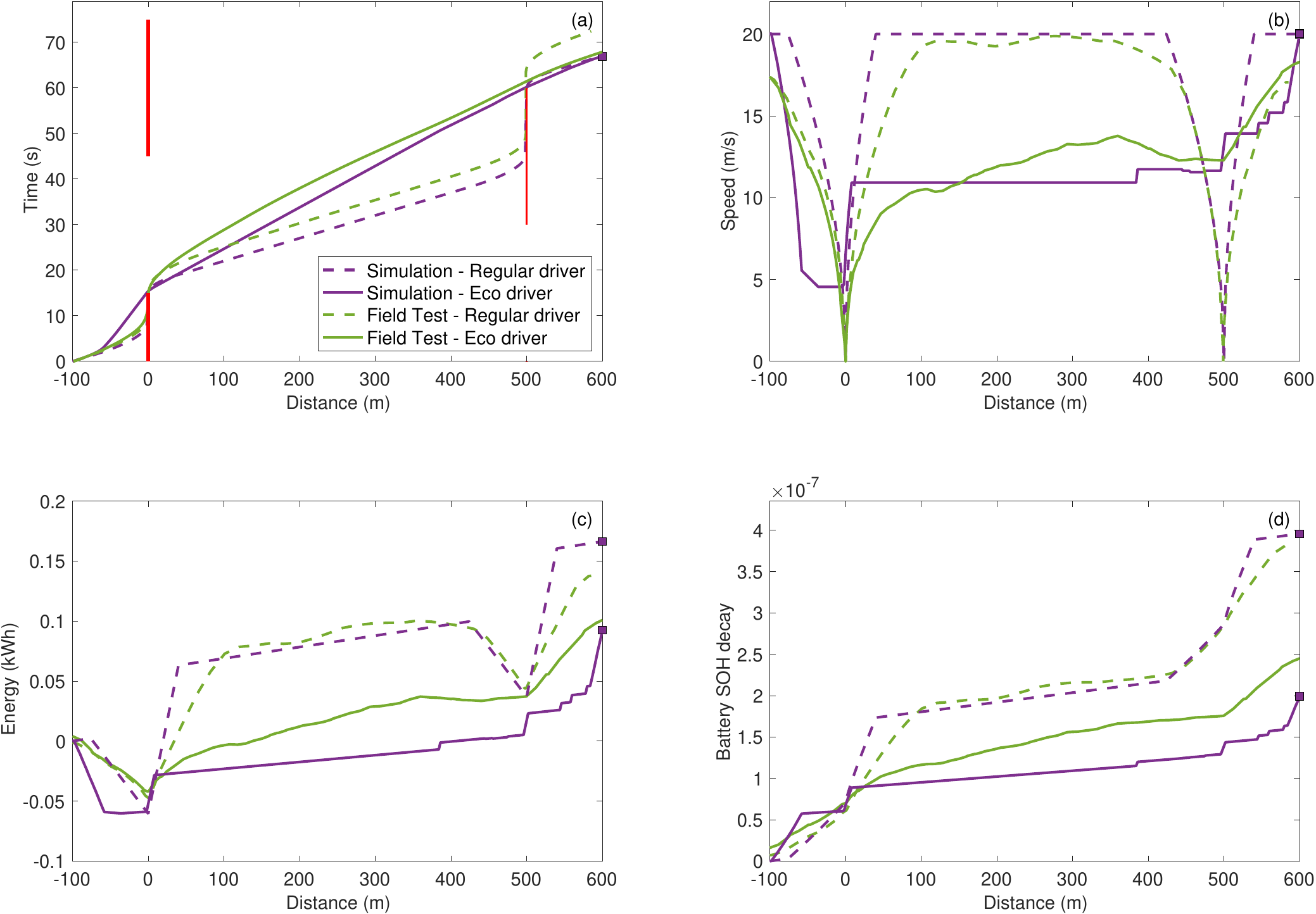}
	\caption{Field test results showing (a) trajectories through time and distance; (b) speed trajectories; (c) energy consumption; and (d) battery capacity decay.}
	\label{fig:itap_trajectories}
\end{figure}

A sample field test result is presented in this section to showcase the benefits of implementing eco-driving with a phone app. A conventional gasoline vehicle was used for the test. Its speed profile was recorded and then plugged into a corresponding EV's numerical model to calculate the electrical energy consumption and battery capacity decay. The trajectory of the vehicle for one test case is shown in \cref{fig:itap_trajectories}(a). The timing of the next two traffic lights is programmed such that the first light turns green at 15 seconds and the second light turns red at 30 seconds after the vehicle enters the test zone. All green/red signal periods are set to 30 seconds. Simulation results are shown in purple and the field test results in green. The test results closely match the simulations except for a little discrepancy right after each light turns green due to human reaction delays. 

The speed profiles are shown in \cref{fig:itap_trajectories}(b). The regular driver lacks timing information and therefore accelerates to the speed limit after the first light and is then forced to brake hard for the second red light. On the other hand, the eco-driver largely followed the app's recommendation to maintain a lower speed after the first light, and only accelerate to the speed limit after the second light turns green. In practice, the driver found it difficult to exactly adhere to the recommended speed particularly at very low speeds around 20 mph, which is why the speed drifted up during the field test. However, this did not change the resulting speed profile drastically.

\begin{figure}
	\centering
	\includegraphics[width=0.9\columnwidth]{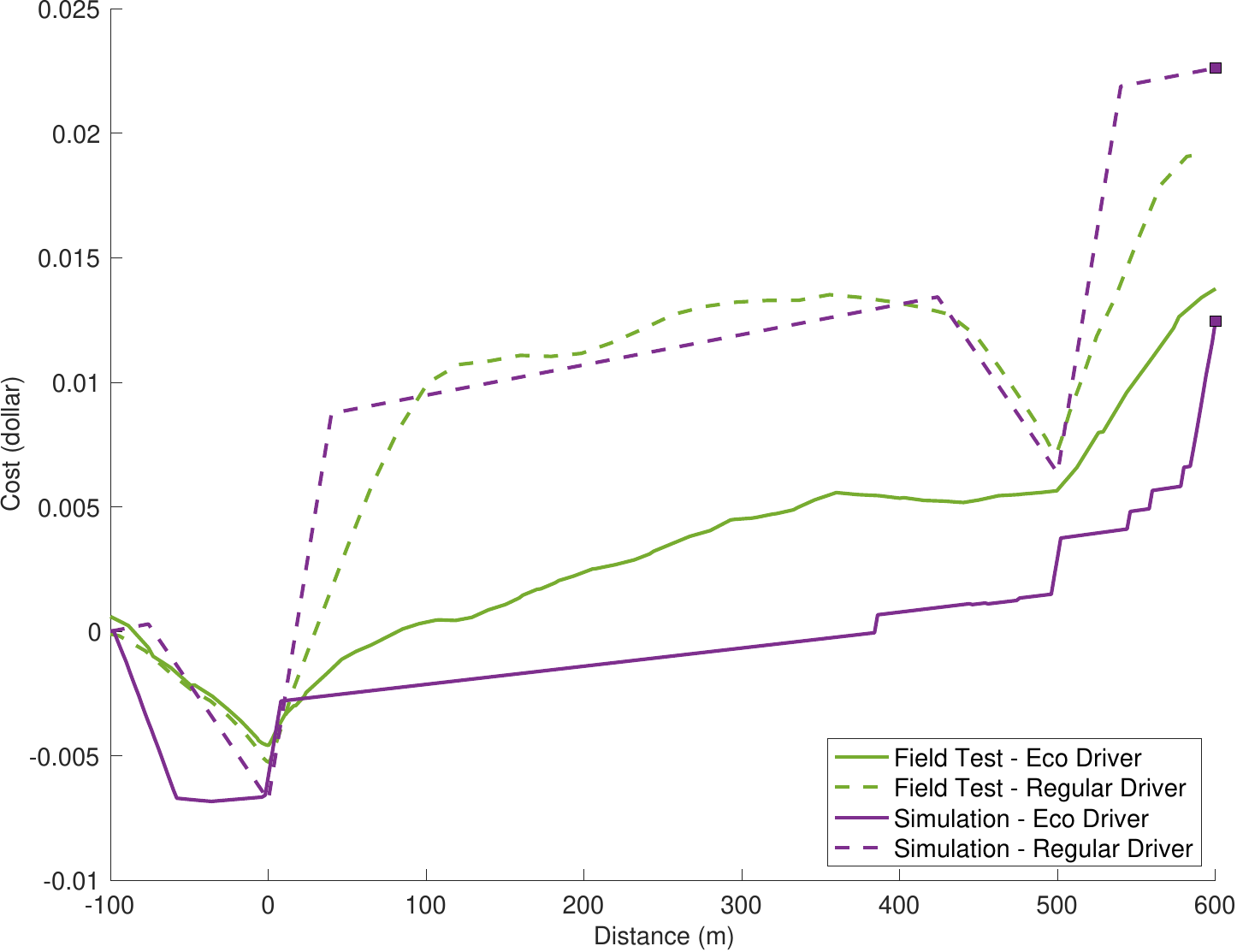}
	\caption{Cost reduction comparison between field test and simulation for the regular and eco-driver.}
	\label{fig:itap_cost}
\end{figure}

The eco-driver's energy saving is around 29\% due to the avoidance of hard braking for the second light as shown in \cref{fig:itap_trajectories}(c). The battery capacity decay is reduced by 38\% for the same reason (\cref{fig:itap_trajectories}(d)). The total saving in dollars is around 32\% (\cref{fig:itap_cost}) when taking into account the cost of electricity and battery replacement. The regular driver in the field test has a lower energy consumption compared to the simulation because the test speed is slightly lower than in the simulation. In contrast, the eco-driver has a higher energy consumption in the field test because he cannot exactly follow the recommendation to maintain a constant low speed. This is also reflected in the battery decay and total cost.

\section{Conclusions}
Dynamic programming was employed to conduct a comprehensive study of the benefits of eco-driving of a connected electric vehicle equipped with V2I communication through two signalized intersections. The cost objective for the DP algorithm included both  energy consumption and battery degradation. Results for the eco-driver are presented and compared with a regular driver. The eco-driving algorithm dramatically reduced the energy consumption through the control zone by decelerating earlier and more gradually while approaching a red light, and cruising at lower speeds while transiting through the traffic signal just as the light turns green. The battery capacity decay and its associated cost is relatively small compared to the energy cost for the baseline battery chemistry (LFP batteries) considered in this study. Higher battery decay rates which are expected for other battery types like NMC and NCA batteries did not significantly affect the optimal speed profiles. Hence the same speed profiles for various battery chemistries could be used in eco-driving to reduce energy consumption and battery decay. Larger battery packs are seen to experience significantly less decay. Thus, the longer lifetimes of larger battery packs also create opportunities to employ these vehicles for grid-integrated functions such as V2G. Finally, a field test was conducted in which a simplified rule-based version of the eco-driving algorithm was implemented as a phone app which issues audio speed recommendations to the driver. The results from field test compared well with the simulations. These convenience of the phone app should facilitate its widespread adoption by all drivers.

\section{Acknowledgment}
This project was funded by the State of Delaware Department of Natural Resources and Environmental Control with cost share provided by the Federal Transit Administration and Delaware Department of Transportation.

\bibliographystyle{IEEEtran}

\end{document}